\documentclass[preprint,proceedings]{rmaa}

\SetYear{2004}
\SetConfTitle{Compact Binaries in the Galaxy and Beyond}

\title{AA Dor --- An Eclipsing Subdwarf -- Brown Dwarf Binary}

\author{
  T. Rauch\altaffilmark{1,2}}

\altaffiltext{1}{Dr.-Remeis-Sternwarte, Sternwartstra\ss e 7, D-96049 Bamberg, Germany
                 (\protect\email{Thomas.Rauch@sternwarte.uni-erlangen.de}).
                }
\altaffiltext{2}{Institut f\"ur Astronomie und Astrophysik, Sand 1, D-72076 T\"ubingen, Germany.
                }

\suppressfulladdresses

\listofauthors{T. Rauch}
\indexauthor{Rauch, T.}

\addkeyword{Stars: binaries: eclipsing}
\addkeyword{Stars: common envelope}
\addkeyword{Stars: evolution}
\addkeyword{Stars: individual: AA Dor}
\addkeyword{Stars: low-mass, brown dwarfs}

\begin{document}
\maketitle 

AA\,Dor is an eclipsing, close, post common-envelope binary (PCEB) consisting of a sdOB primary star 
and an unseen secondary with an extraordinary small mass 
($M_2 \approx 0.066 \mathrm{M}_\odot$) -- formally a brown dwarf (see Rauch 2004 for details). 
In a spectral analysis of AA\,Dor, Rauch (2000) determined
$T_\mathrm{eff} = 42$\,kK and $\log g = 5.2$ (cgs). 
The determination of the components' masses by comparison of these results with evolutionary models
of Driebe et al\@. (1998) shows a discrepancy to masses derived from radial-velocity and the eclipse curves
(Hilditch et al\@. 2003) --
$\log g = 5.5$ would be necessary in order to achieve an intersection at $M_1 \approx 0.330 \mathrm{M}_\odot$. 

Possible reasons for this discrepancy may be too optimistic error ranges in
Rauch (2000) or in the analysis of light curve and radial-velocity curve, or that the evolutionary models
of Driebe et al\@. (1998) are not appropriate in the case of AA\,Dor since these
are post-RGB models for non-CE stars. 

Since the decrement of the hydrogen Balmer series is a sensitive indicator for $\log g$, 
107 high-resolution \'echelle spectra with short exposure times (180 sec) 
have been taken in Jan 2001 with UVES (UV-visual \'echelle spectrograph) attached to the ESO VLT.
Additional medium-resolution longslit spectra have been taken at the 2.3m telescope at SSO in Sept 2003
with the DBS (double beam spectrograph). However, the analysis of both, the UVES spectra
(Rauch \& Werner 2003) and the DBS spectra, shows that a $\log g$ higher than 5.2 results in a worse fit
to the observation. 

\begin{figure}[!t]
  \includegraphics[width=\columnwidth]{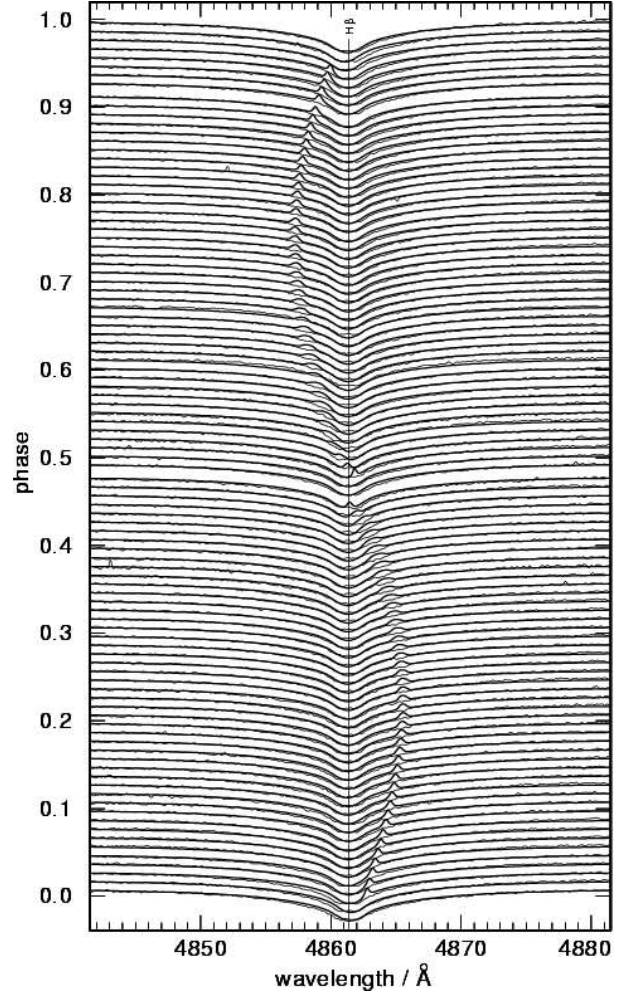}
  \caption{Section of the UVES spectra around H\,$\beta$ compared with
synthetic spectra. A weak H\,$\beta$ emission is used to represent
the secondary's radiation. }
  \label{fig:phase}
\end{figure}

Since the secondary is heated by irradiation of the primary up to $\approx 20\,\mathrm{kK}$, 
one can expect a weak H\,$\beta$ emission in the UVES spectra. However, we do not find any signature of the
secondary. The emission in the line core of H\,$\beta$  (Fig.\,\ref{fig:phase}) comes clearly from the primary,
its phase dependence is likely due to an irradiation effect of the heated secondary on the primary
which increases its $T_\mathrm{eff}$ by $\approx 7\,\mathrm{kK}$, i.e\@. the primary -- taken as an isolated star --
would have only $T_\mathrm{eff}\approx 35\,\mathrm{kK}$, resulting in a $\approx 10$\% smaller mass. A phase-dependent
spectral analysis is presently performed in order to investigate on this effect. It appears possible that
this is one of the main reasons for the disagreement in the mass-radius relation described above.

\begin{thebibliography}

\bibitem{Dea98} Driebe, T., Sch\"onberner, D., Bl\"ocker, T., \& Herwig, F. 1998, A\&A, 339, 129
\bibitem{Hea03} Hilditch, R.W., Kilkenny, D., \& Lynas-Gray, A.E., Hill, G. 2003, MNRAS, 344, 644
\bibitem{Ra00}  Rauch, T. 2000, A\&A, 356, 665
\bibitem{Ra04}  Rauch, T. 2004, Ap\&SS, in press
\bibitem{RW03}  Rauch, T., \& Werner, K. 2003, A\&A, 400, 271

\end{thebibliography}
\end{document}